\documentstyle[pra,preprint,aps,epsf]{revtex}

\newcommand{\rme}{{\rm e}}
\newcommand{\rmi}{{\rm i}}
\newcommand{\rmd}{{\rm d}}
\newcommand{\Or}{{\cal O}}
\begin{document}

\draft
\title{Semiclassical dynamics of a spin-$\frac{1}{2}$ in an arbitrary
magnetic field}

\author{Adrian Alscher and Hermann Grabert}

\address{Fakult\"at f\"ur Physik, Albert-Ludwigs-Universit\"at,\\
Hermann-Herder-Strasse 3, D-79104 Freiburg, Germany}
\date{\today}
\maketitle 
\tighten
\begin{abstract}
The spin coherent state path integral describing the dynamics of a
spin-$\frac{1}{2}$-system in a magnetic field of arbitrary
time-dependence is considered. Defining the path integral as the limit
of a Wiener regularized expression, the semiclassical approximation
leads to a continuous minimal action path with jumps at the
endpoints. The resulting semiclassical propagator is shown to coincide
with the exact quantum mechanical propagator. A non-linear
transformation  of the angle variables   allows for a
determination of the semiclassical path and the jumps without solving
a boundary-value problem. The semiclassical spin dynamics is thus
readily amenable to numerical methods.
\end{abstract}


 \narrowtext

\section{Introduction}
The path integral representation of a quantum system is 
very helpful to visualize the quantum dynamics in terms of classical 
concepts. In particular, spin coherent states allow for a
representation of a spin as a point on a unit sphere depicting the
dynamics in terms of pseudo-classical spin rotations. Unfortunately, in
the standard spin coherent state path integral, the action contains no
terms quadratic in the velocities, and the typical paths are therefore
not continuous, as it is the case with the familiar Feynman
configuration space path integral. This problem has attracted
considerable attention of mathematical physicists
\cite{klauder}-\cite{bodmann}. In most of these studies a spin in a
constant magnetic field is considered which allows for an explicit
solution. Other work \cite{ellinas}-\cite{cabra}  allowing for
time-dependent fields examines the discrete time-lattice version of
the path integral, and it is usually concluded \cite{ercolessi,shibata} that
only formal calculations are possible once the continuous path
integral is employed.

Here we re-examine the semi\-classical propagator of the continuous
time path integral starting from \mbox{Klauder's}
obser\-vation \cite{klauder} that a Wiener regularized coherent state path 
integral for the free spin, that is a spin in the absence of a magnetic
field, allows for a well-defined stationary phase
approximation that turns out to be exact. We will show that
essentially the same type of semiclassical approximation leads to the
exact propagator also in the presence of a magnetic field of arbitrary
time dependence. 

The paper is organized as follows. In section~2 we introduce the basic
notation and the spin coherent state path integral for a
spin-$\frac{1}{2}$-system. We then present, in section~3, 
a spherical Wiener measure regularizing the path integral  
and discuss the semiclassical approximation which is shown to
become exact. In section~4 we transform the angle variables to
variables that allow for a more effective calculation of semiclassical
propagators and calculate the spin coherent propagators for two
models. Finally, in section~5, we present our conclusions.

\section{Spin coherent state path integral}
We consider a spin-$\frac{1}{2}$ described by the spin operators
$S_{i},\, (i=x,y,z)$ with the two-dimensional Hilbert space spanned,
e.g., by the eigenvectors $\bigl|\uparrow\bigr>$ and 
$\bigl|\downarrow\bigr>$ of $S_z$. For each orientation in real space
characterized by a polar angle $\vartheta$ and an azimuthal angle
$\varphi$  we may introduce a spin coherent state 
\cite{perelomov} (we put $\hbar=1$)
\begin{equation}
\bigl|\Omega\bigr>  \equiv
\bigl|\vartheta\varphi\bigr> = \rme^{-\rmi\varphi S_{z}}
\rme^{-\rmi \vartheta S_{y}} \bigl|\uparrow\bigr>.
\label{eq20} 
\end{equation}
These states are not orthogonal but form an overcomplete basis in the
Hilbert space. The overlap of two coherent states reads
\begin{equation}
\left<\Omega''|\Omega'\right>=
\cos\left(\frac{\vartheta''}{2}\right)\cos\left(\frac{\vartheta'}{2}\right)
\rme^{\frac{\rmi}{2}(\varphi''-\varphi')} + 
\sin\left(\frac{\vartheta''}{2}\right)
\sin\left(\frac{\vartheta'}{2}\right)
\rme^{-\frac{\rmi}{2}(\varphi''-\varphi')},
\label{eq21}
\end{equation}
and the identity may be represented as
\begin{equation}
I =\frac{1}{2\pi}\int \rmd\cos(\vartheta) \rmd\varphi 
\left|\Omega\right>\left<\Omega\right|.
\label{eq22b}  
\end{equation}
Furthermore, the matrix elements of the spin operators take the form
\begin{eqnarray}
 \left<\Omega''|S_x|\Omega'\right> &= \frac{1}{2}\left[
 \cos\left(\frac{\vartheta''}{2}\right)\sin\left(\frac{\vartheta'}{2}\right)
\rme^{\frac{\rmi}{2}(\varphi''+\varphi')} + 
 \sin\left(\frac{\vartheta''}{2}\right)\cos\left(\frac{\vartheta'}{2}\right)
\rme^{-\frac{\rmi}{2}(\varphi''+\varphi')} \right]
 \nonumber\\
    \left<\Omega''|S_y|\Omega'\right> &= \frac{1}{2\rmi}\left[
 \cos\left(\frac{\vartheta''}{2}\right)\sin\left(\frac{\vartheta'}{2}\right)
\rme^{\frac{\rmi}{2}(\varphi''+\varphi')} - 
 \sin\left(\frac{\vartheta''}{2}\right)\cos\left(\frac{\vartheta'}{2}\right)
\rme^{-\frac{\rmi}{2}(\varphi''+\varphi')} \right]
 \nonumber\\
  \left<\Omega''|S_z|\Omega'\right> &= \frac{1}{2}\left[
 \cos\left(\frac{\vartheta''}{2}\right)\cos\left(\frac{\vartheta'}{2}\right)
\rme^{\frac{\rmi}{2}(\varphi''-\varphi')} - 
 \sin\left(\frac{\vartheta''}{2}\right)\sin\left(\frac{\vartheta'}{2}\right)
\rme^{-\frac{\rmi}{2}(\varphi''-\varphi')} \right].
\label{eq22}
\end{eqnarray}
%
%
%
%
Let us consider  a spin in a magnetic field of arbitrary time
dependence described by the Hamiltonian
\begin{equation}
H(t)=B_x(t)S_x+B_y(t)S_y+B_z(t)S_z, 
\label{eq22c}
\end{equation}
which gives rise to the unitary time evolution operator
\begin{equation}
U(t)={\cal T}_s\exp\left\{-\rmi\int_0^{t}\rmd s\,H(s)\right\},
\label{eq22d}
\end{equation}
where ${\cal T}_s$ is the time ordering operator. $U(t)$ can be shown
to be of the form \cite{gilmore}
\begin{equation}
U(t)=\left(\begin{array}{cc}
a(t) & b(t) \\  -b^{*}(t) & a^{*}(t) \\
\end{array}\right),\qquad|a(t)|^2+|b(t)|^2=1,
\label{eq21b}
\end{equation}
where the coefficients obey the linear differential equations
\begin{eqnarray}
\dot a(t)&= -\frac{\rmi}{2}B_z(t)a(t) 
+\frac{1}{2}\left[\rmi B_x(t)+B_y(t)\right]b^{*}(t)\nonumber\\
\dot b(t)&= -\frac{\rmi}{2}B_z(s)b(t) 
-\frac{1}{2}\left[\rmi B_x(t)+B_y(t)\right]a^{*}(t).
\label{eq21c}
\end{eqnarray}
Employing a Trotter decomposition, the propagator may be written as 
\begin{equation}
  \bigl<\Omega''\bigr|U(t)\bigl|\Omega'\bigr>=
\lim_{\epsilon\to 0} 
\int\prod_{k=1}^{n}\frac{\rmd\cos(\vartheta_k) 
\rmd\varphi_k}{2\pi} \prod_{k=0}^{n}
\bigl<\Omega_{k+1}\bigr|{\cal
T}_s\exp\{-\rmi\int_{k\epsilon}^{(k+1)\epsilon}
\rmd s\, H(s)\}\bigl|\Omega_{k}\bigr>,
\label{eq23} 
\end{equation}
where $\epsilon=t/n$, $\Omega_{0}=\Omega'$,
$\Omega_{n+1}=\Omega''$. Now, for $\epsilon\rightarrow 0$ we have
\begin{eqnarray}
 \bigl<\Omega_{k+1}\bigr|{\cal
T}_s\exp\{-\rmi\int_{k\epsilon}^{(k+1)\epsilon}\rmd s\, H(s)\}
\bigl|\Omega_{k}\bigr>
=&\bigl<\Omega_{k+1}\bigr.\bigl|\Omega_{k}\bigr>
\left(1-\rmi\epsilon\frac{\bigl<\Omega_{k+1}\bigr|H(k\epsilon)
\bigl|\Omega_{k}\bigr>}
{\bigl<\Omega_{k+1}\bigr.\bigl|\Omega_{k}\bigr>} \right) \nonumber\\
&+ \Or(\epsilon^2), \label{eq24} 
\end{eqnarray}
and the right hand side of equation (\ref{eq23}) can be expressed as
\begin{displaymath}
\lim_{\epsilon\to 0} \int\prod_{k=1}^{n}\frac{\rmd\cos(\vartheta_k)
 \rmd\varphi_k}{2\pi} \exp\left\{\sum_{k=0}^{n}
\left[\log{\bigl<\Omega_{k+1}\bigr.\bigl|\Omega_{k}\bigr>}-
\rmi\epsilon\frac{\bigl<\Omega_{k+1}\bigr|H(k\epsilon)
\bigl|\Omega_{k}\bigr>}
{\bigl<\Omega_{k+1}\bigr.\bigl|\Omega_{k}\bigr>} \right]\right\}.
\end{displaymath}
With the assumption that for $\epsilon\rightarrow 0$ the paths
$\Omega(s)$ remain continuous, we may expand the terms in the exponent
as 
\begin{eqnarray}
\log{\bigl<\Omega_{k+1}\bigr.\bigl|\Omega_{k}\bigr>}&=&
\frac{\rmi}{2}\frac{\cos(\vartheta_{k+1})+\cos(\vartheta_k)}{2}
(\varphi_{k+1}-\varphi_{k})\label{eq26}\\
& &-\frac{1}{8}\left[(\vartheta_{k+1}-\vartheta_k)^2 
+\sin^2(\vartheta_k)(\varphi_{k+1}-\varphi_k)^2 \right]
+\Or(\delta \Omega^3), \nonumber
\end{eqnarray}
and
\begin{equation}
\epsilon\frac{\bigl<\Omega_{k+1}\bigr|H(k\epsilon)\bigl|\Omega_{k}\bigr>} 
{\bigl<\Omega_{k+1}\bigr.\bigl|\Omega_{k}\bigr>} =
\epsilon\bigl<\Omega_{k}\bigr|H(k\epsilon)\bigl|\Omega_{k}\bigr>
+\Or(\epsilon\delta \Omega) \label{eq28}.
\end{equation}
While the term of order $\delta \Omega^2$ in equation (\ref{eq26}) has the
form of the line element on the sphere, this term does not lead 
to a Wiener measure in the path integral, since there are no factors
of $\epsilon$ in the denominator. Hence, the assumption of continuous 
paths is obsolete, and the resulting continuous path integral 
\begin{equation}
 \bigl<\Omega''\bigr|U(t)\bigl|\Omega'\bigr>=
\int_{(\vartheta',\varphi')}^{(\vartheta'',\varphi'')}
{\cal D}\cos(\vartheta)\,{\cal D}\varphi 
\exp\left\{ \rmi\int_{0}^{t}\rmd s\left[\frac{1}{2}
\cos(\vartheta)\dot\varphi-H(\vartheta,\varphi,s)\right]  \right\},
\label{eq28b}
\end{equation}
where
\begin{eqnarray}
 H(\vartheta,\varphi,t)&=&\bigl<\Omega\bigr|H(t)\bigl|\Omega\bigr>
\label{eq28c}\\
&=&\frac{1}{2}\left[B_x(t)\sin(\vartheta)\cos(\varphi)+
B_y(t)\sin(\vartheta)\sin(\varphi)+B_z(t)\cos(\vartheta)\right], 
\nonumber
\end{eqnarray}
has only formal meaning.

\section{Wiener regularization and semiclassical approximation}
Following Klauder \cite{klauder} the ill-defined path integral
(\ref{eq28b}) can be turned into a meaningful expression if the
propagator is written as
\begin{equation}
\bigl<\Omega''\bigr|U(t)\bigl|\Omega'\bigr>=\lim_{\nu\to\infty} 
\int \rmd \mu_{\rm{W}}\,\exp{ \biggl\{ \rmi\int_{0}^{t} \rmd s 
\Bigl[ \frac{1}{2}\cos(\vartheta)\dot\varphi - 
H(\vartheta,\varphi,s) \Bigr]\biggr\} } ,
\label{eq29}
\end{equation}
where
\begin{equation}
\rmd\mu_{\rm{W}}=N\prod_{s=0}^{t} \rmd\cos(\vartheta(s)) 
\rmd\varphi(s) \,\exp{ \biggl\{
- \frac{1}{4\nu}\int_{0}^{t} ds \Bigl[\dot\vartheta^2
+\sin^2(\vartheta)\dot\varphi^2 \Bigr]\biggr\} }
\label{eq30}
\end{equation}
is a Wiener measure on the unit sphere which enforces that only
continuous Brownian motion paths contribute to the path integral. This
amounts to replacing the action of the spin by
\begin{equation}
S_{\nu}[\Omega(s)]= \int_{0}^{t} \rmd s
\left\{ \frac{\rmi}{4\nu} 
\left[ \dot\vartheta^2+\sin^2(\vartheta)\dot\varphi^2\right] 
+ \frac{1}{2}\cos(\vartheta)\dot\varphi - H(\vartheta,\varphi,s) \right\}.
\label{eq29b}
\end{equation}
In the limit $\nu\rightarrow\infty$, the $\nu$-dependent terms in the
action vanish, and formally the previous expression (\ref{eq28b}) is
recovered. 

Let us now investigate the semiclassical approximation of the path
integral. For finite $\nu$ the Euler-Lagrange equations following from
the action (\ref{eq29b}) read
\begin{eqnarray}
\frac{1}{2}\sin(\vartheta)\dot{\varphi}+\frac{\partial
H}{\partial\vartheta} &=&
-\frac{\rmi}{2\nu} \left[\ddot\vartheta - 
\sin(\vartheta)\cos(\vartheta)\dot\varphi^2\right]
\nonumber\\
\frac{1}{2}\sin(\vartheta)\dot\vartheta
-\frac{\partial H}{\partial\varphi}&=&
\frac{\rmi}{2\nu}\left[ \sin^2(\vartheta)\ddot\varphi
+2\sin(\vartheta)\cos(\vartheta)\dot\vartheta\dot\varphi
\right].\label{eq32}
\end{eqnarray}
For given boundary conditions
$\Omega(0)=\Omega'\equiv (\vartheta',\varphi')$,
$\Omega(t)=\Omega''\equiv (\vartheta'',\varphi'')$ and $t\gg1/ \nu$, these
equations have for small and intermediate times $s \,(s\ll t-1/ \nu)$ a
solution of the form
\begin{equation}
\cos(\vartheta(s))=\cos(\bar\vartheta(s))
+\left[\cos(\vartheta')-\cos(\bar \vartheta')\right]\rme^{-\nu s},
\label{eq33a}
\end{equation}
and
\begin{eqnarray}
\varphi(s)&=& \bar \varphi(s)+ \varphi' - \bar \varphi'
+\frac{\rmi}{2}\log\left[\frac{1+\cos(\vartheta')}{1-\cos(\vartheta')}\right]
\nonumber\\
&&-\frac{\rmi}{2}\log\left[\frac
{1+\cos(\bar \vartheta')+(\cos(\vartheta')-
\cos(\bar \vartheta'))\rme^{-\nu s}}
{1-\cos(\bar \vartheta')-(\cos(\vartheta')-
\cos(\bar\vartheta'))\rme^{-\nu s}}\right], 
\label{eq34}
\end{eqnarray}
while for intermediate and large times $s \,(s\gg1/ \nu)$ the solution becomes 
\begin{equation}
\cos(\vartheta(s))=\cos(\bar\vartheta(s))
+\left[\cos(\vartheta'')-\cos(\bar \vartheta'')\right]\rme^{-\nu(t-s)},
\label{eq33b}
\end{equation}
and
\begin{eqnarray} 
\varphi(s)&=&\bar \varphi(s)+ \varphi''  - \bar \varphi''
-\frac{\rmi}{2}\log\left[\frac{1+\cos(\vartheta'')}{1-\cos(\vartheta'')}\right]
\nonumber\\
&& +\frac{\rmi}{2}\log\left[\frac
{1+\cos(\bar \vartheta'')+(\cos(\vartheta'')
-\cos(\bar \vartheta''))\rme^{-\nu(t-s)}}
{1-\cos(\bar \vartheta'')-(\cos(\vartheta'')
-\cos(\bar \vartheta''))\rme^{-\nu(t-s)}}\right]. 
\label{eq35}
\end{eqnarray}
Here, $\bar\Omega(s)\equiv(\bar\vartheta(s),\bar\varphi(s))$ is a solution
of the classical equations of motion
\begin{eqnarray}
\frac{1}{2}\sin(\bar\vartheta)\dot{\bar\varphi}&=&
-\frac{\partial H}{\partial\bar\vartheta}
\nonumber \\
\frac{1}{2}\sin(\bar\vartheta)\dot{\bar\vartheta}&=&
\frac{\partial H}{\partial\bar\varphi}, 
\label{eq45} 
\end{eqnarray}
with the boundary conditions $\bar\Omega(0)
=\bar\Omega'\equiv(\bar\vartheta',\bar\varphi')$ and $\bar
\Omega(t)=\bar\Omega''\equiv(\bar\vartheta'',\bar\varphi'')$. Note that the
solution (\ref{eq33a}) and (\ref{eq34}) describes a jump within the time
interval $1/ \nu$ from the initial state $\Omega'$ to the starting
point $\bar\Omega'$ of the classical trajectory
(\ref{eq45}). Likewise, for $s$ near $t$, 
the solution (\ref{eq33b}) and (\ref{eq35}) describes a jump from the  
endpoint $\bar\Omega''$ of the classical trajectory to the final state
$\Omega''$. Now, in order that the short time and
the long time solutions coincide for intermediate times $1/
\nu\ll s \ll t-1/ \nu$, the boundary conditions of the classical path must
obey the relations
\begin{eqnarray}
\tan\left(\frac{\bar\vartheta'}{2}\right)\rme^{\rmi\bar\varphi'} 
&=& \tan\left(\frac{\vartheta'}{2}\right)\rme^{\rmi\varphi'}
\label{eq50}\\
\tan\left(\frac{\bar\vartheta''}{2}\right)\rme^{-\rmi\bar\varphi''} 
&=&\tan\left(\frac{\vartheta''}{2}\right)\rme^{-\rmi\varphi''}.
\label{eq55}
\end{eqnarray}
This determines the size of the jumps of the semiclassical trajectory
near the endpoints.

Inserting the semiclassical path (\ref{eq33a})-(\ref{eq35}) into the
action $S_{\nu}$ and taking the limit $\nu\rightarrow\infty$ one finds
\begin{equation}
  \exp\left\{\rmi S_{\rm{cl}}\left[\Omega(s)\right]\right\}=
\sqrt{\frac{\sin(\vartheta')\sin(\vartheta'')}
{\sin(\bar\vartheta')\sin(\bar\vartheta'')} }
\exp\left\{\rmi\int_{0}^{t}\rmd s
\left[\frac{1}{2}\cos(\bar\vartheta)\dot{\bar{\varphi}} 
-H(\bar \vartheta,\bar\varphi,s)\right]\right\}.
\label{eq65} 
\end{equation}
Klauder has shown that for $H(\Omega)=0$ the expression (\ref{eq65})
coincides  with the overlap
$\bigl<\Omega''\bigr.\bigl|\Omega'\bigr>$, so that this ``dominant
stationary phase approximation'' \cite{klauder} without fluctuations
becomes exact for a free spin. In general, for a non-vanishing
Hamiltonian, Klauder has concluded that (\ref{eq65}) ``cannot be
expected to provide the correct result by itself''. In fact, in later
work \cite{daubechies} he has suggested a different definition of the
spin coherent path integral. However, we will prove now that for
any Hamiltonian $H(t)$ the exact propagator is given by
\begin{equation}
\bigl<\Omega''\bigr|U(t)\bigl|\Omega'\bigr>=
\exp\left\{\rmi S_{\rm{cl}}\left[\Omega(s)\right]\right\}.
\label{eq66x}
\end{equation}

First we rewrite the overlap between the initial state and the 
starting point of the classical trajectory as
\begin{eqnarray}
\bigl<\bar\Omega'\bigr.\bigl|\Omega'\bigr>&=&
\cos\left(\frac{\bar\vartheta'}{2}\right)\cos\left(\frac{\vartheta'}{2}\right)
\rme^{\frac{\rmi}{2}(\bar\varphi'-\varphi')} + 
\sin\left(\frac{\bar\vartheta'}{2}\right)\sin\left(\frac{\vartheta'}{2}\right)
\rme^{-\frac{\rmi}{2}(\bar\varphi'-\varphi')}
\nonumber\\
&=&\frac{
\sqrt{\sin(\vartheta')\sin(\bar\vartheta')}
\left[1+ \tan\left(\frac{\bar\vartheta'}{2}\right)
\tan\left(\frac{\vartheta'}{2}\right)\rme^{-\rmi(\bar\varphi'-\varphi')} \right]}
{\sqrt{4\tan\left(\frac{\bar\vartheta'}{2}\right)
\tan\left(\frac{\vartheta'}{2}\right)\rme^{-\rmi(\bar\varphi'-\varphi')}}}.
\label{eq66a} 
\end{eqnarray}
Making use of the jump condition (\ref{eq50}), this overlap can be
expressed as
\begin{equation}
\bigl<\bar\Omega'\bigr.\bigl|\Omega'\bigr>=
\sqrt{\frac{\sin(\vartheta')}{\sin(\bar\vartheta')}}.
\label{eq66b} 
\end{equation}	
Likewise, from (\ref{eq55}) we find for the jump at the endpoint 
\begin{equation}
\bigl<\Omega''\bigr.\bigl|\bar\Omega''\bigr>
=\sqrt{\frac{\sin(\vartheta'')}{\sin(\bar\vartheta'')}}.
\label{eq66c} 
\end{equation}	
Therefore, we have from equation (\ref{eq65})
\begin{equation}
  \exp\left\{\rmi S_{\rm{cl}}\left[\Omega(s)\right]\right\}=
\bigl<\Omega''\bigr.\bigl|\bar\Omega''\bigr>
\exp\left\{\rmi\int_{0}^{t}\rmd s
\left[\frac{1}{2}\cos(\bar\vartheta)\dot{\bar{\varphi}} 
-H(\bar \vartheta,\bar\varphi,s)\right]\right\}
\bigl<\bar\Omega'\bigr.\bigl|\Omega'\bigr>.
\label{eq66} 
\end{equation}
Now, the time evolution operator (\ref{eq21b}) acts  on 
a coherent state (\ref{eq20}) as
\begin{eqnarray}
U(t)\bigl|\Omega\bigr>&=&\left[ \cos\left(\frac{\vartheta}{2}\right)
\rme^{-\frac{\rmi}{2}\varphi}a(t)+\sin\left(\frac{\vartheta}{2}\right)
\rme^{\frac{\rmi}{2}\varphi}b(t)\right]\Bigl|\uparrow\Bigr>\nonumber\\
& & + 
\left[-\cos\left(\frac{\vartheta}{2}\right)
\rme^{-\frac{\rmi}{2}\varphi}b^{*}(t)+\sin\left(\frac{\vartheta}{2}\right)
\rme^{\frac{\rmi}{2}\varphi}a^{*}(t)\right]\Bigl|\downarrow\Bigr>.
\label{su0n}
\end{eqnarray}
Apart from a phase factor, the right hand side is again a spin coherent state of
the form (\ref{eq20}). Hence, 
\begin{equation}
U(t)\bigl|\Omega\bigr>=\exp\left\{\rmi\Phi(t)\right\}\Bigl|\Omega(t)\Bigr>,
\label{su1} 
\end{equation}
where $\Omega(t)$ follows from equation (\ref{su0n}) as 
\begin{eqnarray}
\vartheta(t)&=&
\arccos\biggl\{\Bigl[|a(t)|^2-|b(t)|^2\Bigr]\cos(\vartheta)\biggr.\nonumber\\
& &+ \biggl. 
\Bigl[a^{*}(t)b(t)\rme^{\rmi\varphi}+
b^{*}(t)a(t)\rme^{-\rmi\varphi}\Bigr]\sin(\vartheta)\biggr\},
\label{su6}
\end{eqnarray}
and
\begin{eqnarray}
\varphi(t)
 &=&-\frac{\rmi}{2}\log\left\{
\frac{a^{*}(t)b^{*}(t) - 2\left[a^{*}(t)^2 \rme^{\rmi\varphi} -
 b^{*}(t)^2 \rme^{-\rmi\varphi}\right]\tan(\vartheta)} 
{a(t)b(t) - 2\left[a(t)^2 \rme^{-\rmi\varphi} - 
b(t)^2\rme^{\rmi\varphi}\right]\tan(\vartheta) } \right\},
\label{su7}
\end{eqnarray}
and where the phase takes the form 
\begin{equation}
\Phi(t)=\frac{1}{2}\varphi(t)-\frac{\rmi}{2}
\log\left[\frac{a(t)\cos\left(\frac{\vartheta}{2}\right)
\rme^{-\rmi\varphi} + b(t)\sin\left(\frac{\vartheta}{2}\right) }
{a^{*}(t)\cos\left(\frac{\vartheta}{2}\right)+
b^{*}(t)\sin\left(\frac{\vartheta}{2}\right)\rme^{-\rmi\varphi}
}\right] .
\label{su5} 
\end{equation}
In this way we obtain in the spin coherent representation
\begin{eqnarray}
 \bigl<\Omega''\bigr|U(t)\bigl|\Omega'\bigr>&=&
a(t)\cos\left(\frac{\vartheta''}{2}\right)\cos\left(\frac{\vartheta'}{2}\right)
\rme^{\frac{\rmi}{2}(\varphi''-\varphi')}+
a^{*}(t)\sin\left(\frac{\vartheta''}{2}\right)\sin\left(\frac{\vartheta'}{2}\right)
\rme^{-\frac{\rmi}{2}(\varphi''-\varphi')}
\nonumber \\ 
&& + b(t)\cos\left(\frac{\vartheta''}{2}\right)\sin\left(\frac{\vartheta'}{2}\right)
\rme^{\frac{\rmi}{2}(\varphi''+\varphi')}
-b^{*}(t)\sin\left(\frac{\vartheta''}{2}\right)\cos\left(\frac{\vartheta'}{2}\right)
\rme^{-\frac{\rmi}{2}(\varphi''+\varphi')}.
\nonumber\\\label{su8}
\end{eqnarray}
Next, let us  show that $\Omega(t)$ is a solution of the classical 
equations of motion (\ref{eq45}) with initial condition $\Omega(0)=\Omega$.
Inserting equation (\ref{eq21c}) for the time derivatives of the
coefficients $a(t)$ and $b(t)$ we find 
\begin{eqnarray}
\frac{\partial\cos(\vartheta(t))}{\partial t}&=&
\frac{\rmi}{2}B_x(t)\biggl\{2\Bigl[a^*(t)b^*(t)-a(t)b(t)\Bigr]\cos(\vartheta)
\biggr.\nonumber\\
&& \biggl.-\left[a^*(t)^2 +b(t)^2\right]\sin(\vartheta)\rme^{\rmi\varphi}
+\left[a(t)^2 +b^*(t)^2\right]\sin(\vartheta)\rme^{-\rmi\varphi}\biggr\}
\nonumber\\
&&+\frac{1}{2}B_y(t)\biggl\{2\Bigl[a^*(t)b^*(t)+a(t)b(t)\Bigr]\cos(\vartheta)
\biggr.\label{su5b}\\
&&\biggl.-\left[a^*(t)^2 -b(t)^2\right]\sin(\vartheta)\rme^{\rmi\varphi}
-\left[a(t)^2 -b^*(t)^2\right]\sin(\vartheta)\rme^{-\rmi\varphi}
\biggr\}
.\nonumber
\end{eqnarray}
Now, using  equations (\ref{su6}) and (\ref{su7}), the  right hand side
simplifies to give
\begin{equation}
\frac{\partial\cos(\vartheta(t))}{\partial t}=
B_x(t)\sin(\vartheta(t))\sin(\varphi(t))
-B_y(t)\sin(\vartheta(t))\cos(\varphi(t)).
\label{su5c}
\end{equation}
In the same way one derives
\begin{equation}
\frac{\partial\varphi(t)}{\partial t}=
-B_x(t)\frac{\cos(\varphi(t))}{\tan(\vartheta(t))}
-B_y(t)\frac{\sin(\varphi(t))}{\tan(\vartheta(t))}+B_z(t).
\label{su5d}
\end{equation}
The equations (\ref{su5c}) and (\ref{su5d}) are readily shown to coincide
with the equations of motion (\ref{eq45}). 
Hence, the time evolution of the labels
$\Omega(t)$ is purely classical.

The phase $\Phi(t)$ in equation (\ref{su1}) may be expressed in classical
terms as well. In order to do so, let us make use of the Schr\"odinger equation for the
operator $U(t)$.  Since $\frac{\partial}{\partial t}U(t)=-\rmi
H(t)U(t)$, we find from equation (\ref{su1})
\begin{equation}
\Bigl<\Omega(t)\Bigr|\frac{\partial}{\partial t}\Bigl|\Omega(t)\Bigr>
=-\rmi\frac{\partial\Phi(t)}{\partial t} -\rmi\bigl<\Omega\bigr|H(t)\bigl|\Omega\bigr>.
\label{su2}
\end{equation}
This gives
\begin{eqnarray}
\Phi(t) &=&\int_{0}^{t} \rmd s
\Bigl<\Omega(s)\Bigr| \rmi\frac{\partial}{\partial s}-H(s)
\Big|\Omega(s)\Bigr> \nonumber\\
&=&\int_{0}^{t}\rmd s\left[
\frac{1}{2}\cos(\vartheta)\dot{\varphi} -H(\vartheta,\varphi,s)\right],
\label{su3}
\end{eqnarray}
where the   right hand side is just the classical action. Since $\bar
\Omega''=\bar\Omega'(t)$, we have from  equations (\ref{su1}) and (\ref{su3})
\begin{equation}
U(t)\bigl|\bar\Omega'\bigr>=\exp\left\{\rmi\int_{0}^{t}\rmd s
\left[\frac{1}{2}\cos(\bar\vartheta)\dot{\bar{\varphi}} 
-H(\bar \vartheta,\bar\varphi,s)\right]\right\}\bigl|\bar\Omega''\bigr>.
\label{su3b} 
\end{equation}
Now, we  are  in the position to rewrite the semiclassical propagator
(\ref{eq65}). Combining  equations (\ref{eq66b}), (\ref{eq66c}) and (\ref{su3b}) we find
\begin{equation}
\exp\left\{\rmi S_{\rm{cl}}\left[\Omega(s)\right]\right\}=
\bigl<\Omega''\bigr.\bigl|\bar\Omega''\bigr>
\bigl<\bar\Omega''\bigr|U(t)\bigl|\bar\Omega'\bigr>
\bigl<\bar\Omega'\bigr.\bigl|\Omega'\bigr>.
\label{su4} 
\end{equation}
On the other hand,  using equation (\ref{su3b}) one obtains
\begin{equation}
\bigl<\Omega\bigr|U(t)\bigl|\bar\Omega'\bigr>=
\bigl<\Omega\bigr.\bigl|\bar\Omega''\bigr>
\bigl<\bar\Omega''\bigr|U(t)\bigl|\bar\Omega'\bigr>.
\label{su4b} 
\end{equation}
Likewise, with $U(t)=U(-t)^{\dagger}$ one finds
\begin{equation}
\bigl<\bar\Omega''\bigr|U(t)\bigl|\Omega\bigr>=
\bigl<\bar\Omega''\bigr|U(t)\bigl|\bar\Omega'\bigr>
\bigl<\bar\Omega'\bigr.\bigl|\Omega\bigr>,
\label{su4c} 
\end{equation}
and equation (\ref{su4}) finally becomes 
\begin{equation}
\exp\left\{\rmi S_{\rm{cl}}\left[\Omega(s)\right]\right\}
=\bigl<\Omega''\bigr|U(t)\bigl|\Omega'\bigr>.
\label{su9}
\end{equation}
This shows that the dominant stationary phase approximation gives the
exact spin propagator. 

To elucidate this point further, we demonstrate
that the semiclassical propagator obeys the Schr\"odinger
equation. From equation (\ref{eq65}) we find for the time rate of change 
\begin{eqnarray}
 \frac{\partial}{\partial t}
\exp\left\{\rmi S_{\rm{cl}}\left[\Omega(s)\right]\right\}&=&
\frac{1}{2}\Biggl\{  
\frac{\cos(\bar\vartheta')}{\sin^2(\bar\vartheta')}
\frac{\partial\cos(\bar\vartheta')}{\partial t}   +
\frac{\cos(\bar\vartheta'')}{\sin^2(\bar\vartheta'')}
\frac{\partial\cos(\bar\vartheta'')}{\partial t}  \Biggr.
\nonumber\\
&&+\rmi\cos(\bar\vartheta'')
\left.\frac{\partial \bar\varphi(s,t)}{\partial s}\right|_{s=t}
-2\rmi H(\bar\vartheta'',\bar\varphi'',t)\nonumber\\
&&
+\rmi\int_{0}^{t}\rmd s \left[ 
-\sin(\bar\vartheta(s,t))\frac{\partial\bar\vartheta(s,t)}{\partial t}
\frac{\partial\bar\varphi(s,t)}{\partial s} 
+\cos(\bar\vartheta(s,t))\frac{\partial^2\bar\varphi(s,t)}{\partial
t\partial s}\right.\nonumber\\
&&\Biggl.\left.
-2\frac{\partial H}{\partial\bar\vartheta}
\frac{\partial\bar\vartheta(s,t)}{\partial t}
-2\frac{\partial H}{\partial\bar\varphi}
\frac{\partial\bar\varphi(s,t)}{\partial t}
\right] \Biggr\}
\exp\left\{\rmi S_{\rm{cl}}\left[\Omega(s)\right]\right\}.
\label{su10}
\end{eqnarray}
Now, the jump conditions (\ref{eq50}) and (\ref{eq55}) give
\begin{equation}
\cos(\bar\vartheta')=\frac
{\left[1+\cos(\vartheta')\right]\rme^{2\rmi\bar\varphi'}
-\left[1-\cos(\vartheta')\right]\rme^{2\rmi\varphi'}}
{\left[1+\cos(\vartheta')\right]\rme^{2\rmi\bar\varphi'}
+\left[1-\cos(\vartheta')\right]\rme^{2\rmi\varphi'}},
\label{su10b}
\end{equation}
and a similar relation for $\cos(\bar\vartheta'')$.
These relations can be used to re-write the first two terms 
on the  right hand side of equation (\ref{su10}) as
\begin{equation}
\frac{\cos(\bar\vartheta')}{\sin^2(\bar\vartheta')}
\frac{\partial\cos(\bar\vartheta')}{\partial t}
=\rmi\cos(\bar\vartheta')\frac{\partial\bar\varphi'}{\partial t}
\label{su11a}
\end{equation}
and
\begin{equation}
\frac{\cos(\bar\vartheta'')}{\sin^2(\bar\vartheta'')}
\frac{\partial\cos(\bar\vartheta'')}{\partial t}
=-\rmi\cos(\bar\vartheta'')\frac{\partial\bar\varphi''}{\partial t}
\label{su11b}.
\end{equation}
Then, after an integration by parts, equation (\ref{su10}) becomes
\begin{eqnarray}
\frac{\partial}{\partial t}
\exp\left\{\rmi S_{\rm{cl}}\left[\Omega(s)\right]\right\}&=&
\Biggl\{
\rmi\int_{0}^{t}ds \left[
\frac{1}{2}\sin[\bar\vartheta(s,t)]
\frac{\partial\bar\vartheta(s,t)}{\partial s} 
-\frac{\partial H}{\partial\bar\varphi}
\right]\frac{\partial\bar\varphi(s,t)}{\partial t}\Biggr.
\nonumber\\
& & 
-\rmi\int_{0}^{t}ds \left[ 
\frac{1}{2}\sin[\bar\vartheta(s,t)]\frac{\partial\bar\varphi(s,t)}{\partial s} 
+\frac{\partial H}{\partial\bar\vartheta}
\right]\frac{\partial\bar\vartheta(s,t)}{\partial t}
\nonumber\\
& &
\Biggl. -\rmi H(\bar\vartheta'',\bar\varphi'',t) \Biggr\}
\exp\left\{\rmi S_{\rm{cl}}\left[\Omega(s)\right]\right\}.
\label{su12}
\end{eqnarray}
Therefore, with the equations of motions (\ref{eq45}), we obtain the
Schr\"odinger equation 
\begin{equation}
\frac{\partial}{\partial t}
\exp\left\{\rmi S_{\rm{cl}}\left[\Omega(s)\right]\right\}=
-\rmi H(\bar\vartheta'',\bar\varphi'',t)\exp\left\{\rmi S_{\rm{cl}}\left[\Omega(s)\right]\right\}.
\label{su13}
\end{equation}
Note that the matrix element of the Hamiltonian at the endpoint
$\bar\Omega''$ of the classical trajectory generates the time rate of
change of the semiclassical propagator and not the matrix element at
the final state $\Omega''$. The semiclassical propagator may thus be
written as 
\begin{equation}
\exp\left\{\rmi S_{\rm{cl}}\left[\Omega(s)\right]\right\}=
\exp\left\{-\rmi\int_0^{t}\rmd s\,
H(\bar\vartheta''(s),\bar\varphi''(s),s)\right\}
\bigl<\Omega''\bigr|\bigl.\Omega'\bigr>.
\label{su13b}
\end{equation}
To demonstrate that the Schr\"odinger equation (\ref{su13}) generates
the exact quantum dynamics, we start from the equation of motion of $U(t)$
\begin{equation}
\frac{\partial}{\partial t}\bigl<\Omega''\bigr|U(t)\bigl|\Omega'\bigr>
=-\rmi\bigl<\Omega''\bigr|H(t)U(t)\bigl|\Omega'\bigr>.
\label{su13c}
\end{equation}
In view of (\ref{su4b}) we have
\begin{eqnarray}
\bigl<\Omega''\bigr|H(t)U(t)\bigl|\Omega'\bigr>&=&
\bigl<\Omega''\bigr|H(t)\bigl|\bar\Omega''\bigr>
\bigl<\bar\Omega''\bigr| U(t)\bigl|\Omega'\bigr>\nonumber\\
&=&\frac{\bigl<\Omega''\bigr|H(t)\bigl|\bar\Omega''\bigr>}
{\bigl<\Omega''\bigr|\bigl.\bar\Omega''\bigr>}
\bigl<\Omega''\bigr|U(t)\bigl|\Omega'\bigr>,
\label{su13d}
\end{eqnarray}
where the first factor in the second line  can also be written as
\begin{equation}
\frac{\bigl<\Omega''\bigr|H(t)\bigl|\bar\Omega''\bigr>}
{\bigl<\Omega''\bigr.\bigl|\bar\Omega''\bigr>}
= \bigl<\bar\Omega''\bigr|H(t)\bigl|\bar\Omega''\bigr>
=H(\bar\vartheta'',\bar\varphi'',t).
\label{su14}
\end{equation}
To show this, we represent the matrix elements (\ref{eq22}) of the
spin operators in the form
\begin{eqnarray}
\bigl<\Omega''\bigr|S_x\bigl|\bar\Omega''\bigr>&=&\frac{1}{2}
\frac{
 \tan{\frac{\bar\vartheta''}{2}}\rme^{\rmi\bar\varphi''} + 
 \tan{\frac{\vartheta''}{2}}\rme^{-\rmi\varphi''}}
{1 + \tan{\frac{\vartheta''}{2}}
\tan{\frac{\bar\vartheta''}{2}}\rme^{\rmi(\bar\varphi''-\varphi'')}   }
\bigl<\Omega''\bigr.\bigl|\bar\Omega''\bigr>
 \nonumber\\
\bigl<\Omega''\bigl|S_y\bigr|\bar\Omega''\bigr>&=&-\frac{\rmi}{2}\frac{
 \tan{\frac{\bar\vartheta''}{2}}\rme^{\rmi\bar\varphi''} - 
 \tan{\frac{\vartheta''}{2}}\rme^{-\rmi\varphi''} }
{1 + \tan{\frac{\vartheta''}{2}}
\tan{\frac{\bar\vartheta''}{2}}\rme^{\rmi(\bar\varphi''-\varphi'')} }
\bigl<\Omega''\bigr.\bigl|\bar\Omega''\bigr>
 \nonumber\\
\bigl<\Omega''\bigr|S_z\bigl|\bar\Omega''\bigr>&=&\frac{1}{2}\frac{ 
1 - \tan{\frac{\vartheta''}{2}}
\tan{\frac{\vartheta'}{2}}\rme^{\rmi(\bar\varphi''-\varphi'')} }
{1 + \tan{\frac{\vartheta''}{2}}
\tan{\frac{\bar\vartheta''}{2}}\rme^{\rmi(\bar\varphi''-\varphi'')} }
\bigl<\Omega''\bigr.\bigl|\bar\Omega''\bigr>,
\label{su15}
\end{eqnarray}
and insert the jump condition (\ref{eq55}) to yield
\begin{eqnarray}
\bigl<\Omega''\bigr|S_x\bigl|\bar\Omega''\bigr>&=&
\frac{1}{2}\sin{\bar\vartheta''}\cos{\bar\varphi''}   
\bigl<\Omega''\bigr.\bigl|\bar\Omega''\bigr>
 \nonumber\\
\bigl<\Omega''\bigr|S_y\bigl|\bar\Omega''\bigr>&=&
\frac{1}{2}\sin{\bar\vartheta''}\sin{\bar\varphi''} 
\bigl<\Omega''\bigr.\bigl|\bar\Omega''\bigr>
 \nonumber\\
\bigl<\Omega''\bigr|S_z\bigl|\bar\Omega''\bigr>&=&
\frac{1}{2}\cos{\bar\vartheta''}
\bigl<\Omega''\bigr.\bigl|\bar\Omega''\bigr>.
\label{su16}
\end{eqnarray}
Then, from equation (\ref{eq22c}), the relation (\ref{su14}) is readily
shown, and  equations (\ref{su13c}) and (\ref{su13d}) combine again to the 
Schr\"odinger equation (\ref{su13}).

\section{Calculation of semiclassical propagators}
In the semiclassical theory described in the previous section 
the starting and end points $\bar\Omega'$ and $\bar\Omega''$ of the classical
trajectory $\Omega(s)$ need to be determined by solving a boundary value
problem. This requires usually some effort. The same problem arises
for the coherent state propagator of a simple harmonic oscillator
\cite{klauder} and there it is useful to rewrite the propagator in
terms of the complex Glauber variables \cite{weissman}.  
Here, we present a nonlinear transformation of the  angle variables of
the semiclassical spin which allows for an
explicit calculation of the spin coherent state propagator 
by solving only an initial value type problem. Let us
introduce the variables \cite{perelomov}
\begin{eqnarray}
\zeta&=&\tan\left(\frac{\vartheta}{2}\right)\rme^{\rmi\varphi}
\nonumber\\
\eta&=&\tan\left(\frac{\vartheta}{2}\right)\rme^{-\rmi\varphi}.
\label{sp5}
\end{eqnarray}
For real angles $\vartheta$ and $\varphi$ one has
$\eta=\zeta^{*}$, and the transformation corresponds to a
stereographic projection from the south pole of the
unit sphere onto the equatorial plane.
%
%
%
%
However, usually the semiclassical trajectories
(\ref{eq33a})-(\ref{eq35})  become complex and $\zeta$ and $\eta$ are
independent variables. Using the inverse transformation
\begin{eqnarray}
\vartheta&=&\arccos\left[\frac{1-\zeta\eta}{1+\zeta\eta}
\right] \nonumber\\
\varphi&=&\arctan\left[\frac{\zeta-\eta}{\rmi(\zeta+\eta)}
\right],
\label{sp1}
\end{eqnarray}
the Hamiltonian (\ref{eq28c}) takes the form
\begin{equation}
H(\zeta,\eta,t)
=\frac{1}{2}\left[ B_x(t)\frac{\zeta+\eta}{1+\zeta\eta}
-\rmi B_y(t)\frac{\zeta-\eta}{1+\zeta\eta}
+B_z(t)\frac{1-\zeta\eta}{1+\zeta\eta}\right],
\label{sp2}
\end{equation}
and the classical action becomes
\begin{eqnarray}
\exp\left\{\rmi S_{\rm{cl}}\left[\Omega(s)\right]\right\}&=&
\sqrt{\frac{(1+\zeta(0)\eta(0))(1+\zeta(t)\eta(t))}
{(1+\zeta'\eta')(1+\zeta''\eta'')} }
\left(\frac{\zeta'\eta'\zeta''\eta''}
{\zeta(0)\eta(0)\zeta(t)\eta(t)} \right)^{\frac{1}{4}}
\nonumber\\
& &\times
\exp\left\{\int_{0}^{t}\rmd s\left[
\frac{(1-\zeta\eta)(\dot \zeta\eta-\zeta\dot\eta )}{4\zeta\eta(1+\zeta\eta)}
-\rmi H(\zeta,\eta,s)\right]\right\}.
\label{sp3}
\end{eqnarray}
The time integral in the exponent may be rewritten as
\begin{eqnarray}
&&\int_{0}^{t}\rmd s\left[
\frac{(1-\zeta\eta)(\dot \zeta\eta-\zeta\dot\eta )}{4\zeta\eta(1+\zeta\eta)}
- \rmi H(\zeta,\eta,s)\right]\nonumber\\
&&= \rmi\int_{0}^{t}\rmd s\left[
\frac{\rmi}{2}\left(\frac{\dot \zeta\eta-\zeta\dot \eta }{1+\zeta\eta}
-\frac{\partial}{\partial s}\log\left[\frac{\zeta}{\eta}
\right] \right)
-H(\zeta,\eta,s)\right].
\nonumber 
\end{eqnarray}
and the jump conditions (\ref{eq50}) and (\ref{eq55}) transform into
the simple boundary conditions
\begin{eqnarray}
\zeta(0)&=&\zeta'\nonumber\\
\eta(t)&=&\eta''.
\label{sp6} 
\end{eqnarray}
Thus, we obtain from equation (\ref{sp3})
\begin{eqnarray}
\exp\left\{\rmi S_{\rm{cl}}[\Omega(s)]\right\}&=&
\sqrt{\frac{(1+\zeta'\eta(0))(1+\zeta(t)\eta'')}
{(1+\zeta'\eta')(1+\zeta''\eta'')} }
\left(\frac{\zeta'' \eta'}{\zeta' \eta''}\right)^\frac{1}{4}
\nonumber\\ 
&&\times
\exp\left\{ \rmi\int_{0}^{t}\rmd s
\left[\frac{\rmi}{2}\frac{\dot \zeta\eta-\zeta\dot \eta}{1+\zeta\eta}
-H(\zeta,\eta,s)\right]\right\}.
\label{sp7} 
\end{eqnarray}
The classical equations of motion (\ref{eq45}) read in terms of the
new variables
\begin{eqnarray}
\dot \zeta &=&- \rmi(1+\zeta\eta)^2\frac{\partial H}{\partial \eta}
\nonumber\\
\dot \eta &=&\rmi(1+\zeta\eta)^2\frac{\partial H}{\partial \zeta}.
\label{sp8} 
\end{eqnarray}
These equations coincide with the Euler-Lagrange equations of the action
\begin{equation}
S'[\zeta(s),\eta(s)]= \int_{0}^{t}\rmd s\left[
\frac{\rmi}{2} \frac{\dot \zeta\eta-\zeta\dot \eta }{1+\zeta\eta}
-H(\zeta,\eta,s)\right].
\label{sp8b}
\end{equation} 
This action and the associated classical equations of motion have been
studied by several authors \cite{kuratsuji,keski,fukui,funahashi1,ercolessi}. 
It is important to note that the
action (\ref{sp8b}) evaluated along the trajectories solving
equations (\ref{sp8}) with boundary conditions (\ref{sp6}) does not
yield the exact quantum mechanical propagator through a relation of
the form (\ref{eq66x}). 

Using the explicit form  (\ref{sp2}) of the
Hamlitonian the equations (\ref{sp8}) decouple and read explicitly
\begin{eqnarray}
\dot \zeta
&=&-\frac{\rmi}{2}B_x(1-\zeta^2)+\frac{1}{2}B_y(1+\zeta^2)+\rmi B_z\zeta
\nonumber\\
\dot \eta
&=&\frac{\rmi}{2}B_x(1-\eta^2)+\frac{1}{2}B_y(1+\eta^2)-\rmi B_z\eta.
\label{sp9} 
\end{eqnarray}
Since the solution has to satisfy the conditions (\ref{sp6}), we see
that the boundary-value problem is now reduced to an initial- or final-value
problem. Moreover, the two equations of motion are complex conjugate.
Inserting the equations of motion (\ref{sp9}) into the exponent of
equation (\ref{sp7}), the time  integral  simplifies and we finally obtain  
\begin{eqnarray}
&&\exp\left\{\rmi S_{\rm{cl}}[\Omega(s)]\right\}=
\sqrt{\frac{(1+\zeta'\eta(0))(1+\zeta(t)\eta'')}
{(1+\zeta'\eta')(1+\zeta''\eta'')} }
\left(\frac{\zeta'' \eta'}{\zeta'\eta''}\right)^\frac{1}{4}
\nonumber\\
&&\times
\exp\left\{-\frac{\rmi}{4}\int_{0}^{t}\rmd s
\biggl[B_x\Bigl(\zeta+\eta\Bigr)-
\rmi B_y\Bigl(\zeta-\eta\Bigr)+2B_z\biggr]\right\}.
\label{sp10} 
\end{eqnarray}

To illustrate the theory we apply it to two specific models.
As a first example we  treat the propagator of a two-state system with
Hamiltonian 
\begin{equation}
H=\Delta S_x + \epsilon S_z,
\label{ex0}
\end{equation}
which describes a variety of systems. The Hamiltonian 
(\ref{ex0}) corresponds to a spin-$\frac{1}{2}$ in a
time-independent magnetic field $\vec B=(\Delta,0,\epsilon)$.   
With the method presented above, we express  this Hamiltonian as 
\begin{equation}
H(\zeta,\eta)=\frac{\Delta}{2}\frac{\zeta+\eta}{1+\zeta\eta}
+\frac{\epsilon}{2}\frac{1-\zeta\eta}{1+\zeta\eta}.
\label{ex1}
\end{equation}
In accordance with the boundary conditions (\ref{sp6}), the equations of
motion (\ref{sp9}) are solved by
\begin{eqnarray}
\zeta(s)&=&-\frac{\epsilon}{\Delta}
-\rmi\frac{\omega}{\Delta}\tan\left\{\omega \frac{s}{2} 
+\arctan\left[\frac{\rmi(\epsilon+\Delta\zeta')}{\omega}\right]  
\right\}
\nonumber\\
\eta(s)&=&-\frac{\epsilon}{\Delta}
-\rmi\frac{\omega}{\Delta}\tan\left\{\omega\frac{t-s}{2} 
+\arctan\left[\frac{\rmi(\epsilon+\Delta\eta'')}{\omega}\right]  
\right\},
\label{ex2} 
\end{eqnarray}
where $\omega=\sqrt{\Delta^2+\epsilon^2}$. The unspecified boundary
values may be written as 
\begin{equation}
\zeta(t)=\frac{\omega \zeta'\cos\left(\frac{\omega t}{2}\right)
 +\rmi(\epsilon\zeta'-\Delta)\sin\left(\frac{\omega t}{2}\right)}
{\omega\cos\left(\frac{\omega t}{2}\right)
 -\rmi(\Delta\zeta'+\epsilon)\sin\left(\frac{\omega t}{2}\right)},
\label{ex3}
\end{equation}
and\begin{equation}
\eta(0)=\frac{\omega\eta''\cos\left(\frac{\omega t}{2}\right)
 +\rmi(\epsilon\eta''-\Delta)\sin\left(\frac{\omega t}{2}\right)}
{\omega\cos\left(\frac{\omega t}{2}\right)
 -\rmi(\Delta\eta''+\epsilon)\sin\left(\frac{\omega t}{2}\right)}.
\label{ex4}
\end{equation}
Now, the time  integral in equation (\ref{sp10}) can be readily solved
\begin{eqnarray}
&&\exp\left\{ -\frac{\rmi}{2}\int_{0}^{t}\rmd s
\left[\frac{1}{2}\Delta\left(\zeta(s)+\eta(s)\right)+\epsilon\right]\right\}=
\nonumber\\
&& \frac{1}{\omega}\left[\omega
\cos\left(\frac{\omega t}{2}\right)
 -\rmi(\Delta\zeta'+\epsilon)
\sin\left(\frac{\omega t}{2}\right)\right]^{\frac{1}{2}}
\left[\omega
\cos\left(\frac{\omega t}{2}\right)
 -\rmi(\Delta\eta''+\epsilon)\sin\left(\frac{\omega t}{2}\right)\right]^{\frac{1}{2}}.
\label{ex5} 
\end{eqnarray}
Combining these relations we obtain from equation (\ref{sp10})
\begin{eqnarray}
\exp\left\{\rmi S_{\rm{cl}}[\Omega(s)]\right\}
&=&
\frac{\left(\frac{\zeta'' \eta'}{\zeta' \eta''}\right)^\frac{1}{4} }
{\sqrt{(1+\zeta'\eta')(1+\zeta''\eta'')}}\biggl[
\left(1+\zeta'\eta''\right)\cos\left(\frac{\omega t}{2}\right) \biggr.
\nonumber\\
&&-\biggl.   \frac{\rmi\left(\epsilon
-\epsilon\zeta'\eta''+\Delta\zeta'+\Delta\eta''\right)}
{\omega}\sin\left(\frac{\omega t}{2}\right)  \biggr].
\label{ex6}
\end{eqnarray}
With the inverse transformation the semiclassical propagator
takes the form of the  right hand side of equation (\ref{su8}) with
\begin{equation}
a(t)=\cos\left(\frac{\omega t}{2}\right)-
\frac{\rmi\epsilon}{\omega}\sin\left(\frac{\omega t}{2}\right),
\label{ex8} 
\end{equation}
and
\begin{equation}
b(t)=-\frac{\rmi\Delta}{\omega}
\sin\left(\frac{\omega t}{2}\right),
\label{ex9} 
\end{equation}
which coincides with the exact quantum mechanical result.

As a second example, we consider the Landau-Zener problem \cite{shore}
\begin{equation}
H=\omega S_x -\gamma^2t S_z,
\label{ex10a}
\end{equation}
which corresponds to a spin-$\frac{1}{2}$ in the 
time-dependent magnetic field $\vec B=(\omega,0,-\gamma^2 t)$. 
The Hamiltonian  now reads
\begin{equation}
H(\zeta,\eta,t)=\frac{\omega}{2}\frac{\zeta+\eta}{1+\zeta\eta}
-\frac{\gamma^2 t}{2}\frac{1-\zeta\eta}{1+\zeta\eta}.
\label{ex10}
\end{equation}
The equations of motion (\ref{sp9}) are of Riccati form, and for the
present model the transformation  
\begin{equation}
\zeta(s)=\frac{a^{*}(s)\zeta -b^*(s)}{b(s)\zeta+a(s)}
\label{ex10b}
\end{equation}
leads to Weber equations for $a(s)$ and $b(s)$ \cite{whittaker} which
are solved in terms of confluent hypergeometric functions
$\Phi(\alpha,\beta,z)$ \cite{gradshteyn}. Accordingly, the solutions
read 
\begin{eqnarray}
\zeta(s)&=\frac{D(s)\zeta' + C(s)}{B(s)\zeta'+A(s)}\nonumber\\
\eta(s)&=\frac{
\left[A(t)A(s)-C(t)B(s)\right]\eta'' -\left[A(t)B(s)-B(t)A(s)\right] }
{\left[C(t)D(s)-D(t)C(s)\right]\eta''+\left[A(t)D(s)-B(t)C(s)\right]},
\label{ex11}
\end{eqnarray}
where
\begin{eqnarray}
A(s)&=&\Phi\left(-\frac{\rmi}{8}\frac{\omega^2}{\gamma^2},
\frac{1}{2},-\frac{\rmi}{2}\gamma^2 s^2\right)
\nonumber\\
B(s)&=&-\frac{\rmi}{2}\omega s \,
\Phi\left(-\frac{\rmi}{8}\frac{\omega^2}{\gamma^2}+
\frac{1}{2},\frac{3}{2},-\frac{\rmi}{2}\gamma^2 s^2\right)
\nonumber\\
C(s)&=&-\frac{\rmi}{2}\omega s\,
\Phi\left(-\frac{\rmi}{8}\frac{\omega^2}{\gamma^2}+1,
\frac{3}{2},-\frac{\rmi}{2}\gamma^2 s^2\right)
\nonumber\\
D(s)&=&\Phi\left(-\frac{\rmi}{8}\frac{\omega^2}{\gamma^2}+
\frac{1}{2},\frac{1}{2},-\frac{\rmi}{2}\gamma^2 s^2\right).
\label{ex12}
\end{eqnarray}
Now, the unspecified boundary values become
\begin{eqnarray}
\zeta(t)&=\frac{D(t)\zeta' +C(t)}{B(t)\zeta'+A(t)}
\nonumber\\
\eta(0)&=\frac{D(t)\eta''+B(t) }{C(t)\eta''+A(t)}.
\label{ex13}
\end{eqnarray}
To solve the time integral in equation (\ref{sp10}) we make
use of  analytic properties of $\Phi(\alpha,\beta,z)$
\cite{gradshteyn} yielding
\begin{eqnarray}
\frac{\rmd}{\rmd s}A(s)&=&-\frac{\rmi}{2}\omega C(s) 
\nonumber\\
\frac{\rmd}{\rmd s}B(s)&=&-\frac{\rmi}{2}\omega D(s)
\nonumber\\
\frac{\rmd}{\rmd s}C(s)&=&-\frac{\rmi}{2}\omega A(s) -\rmi \gamma^2 s C(s)
\nonumber\\
\frac{\rmd}{\rmd s}D(s)&=&-\frac{\rmi}{2}\omega B(s)-\rmi \gamma^2 s D(s).
\label{ex14}
\end{eqnarray}
We then obtain 
\begin{equation}
\exp\left\{ -\frac{\rmi}{4}\omega\int_{0}^{t}\rmd s 
\left[\zeta(s)+\eta(s)\right]\right\}=
\sqrt{B(t)\zeta'+A(s)}\sqrt{C(t)\eta''+D(t)}.
\label{ex15} 
\end{equation}
Inserting the results into (\ref{sp10}) and expressing the
final and initial states again in terms of angles, the semiclassical
propagator takes the form (\ref{su8}) where
\begin{equation}
a(t)=\exp\left\{\frac{\rmi}{4} \gamma^2 t^2\right\}A(t)
\label{ex17} 
\end{equation}
and
\begin{equation}
b(t)=\exp\left\{\frac{\rmi}{4} \gamma^2 t^2\right\}B(t).
\label{ex18}
\end{equation}
This is again the exact quantum mechanical result.

\section{Conclusions}
We have analyzed the spin coherent state path integral for a
spin-$\frac{1}{2}$ in a magnetic field of arbitrary time
dependence. To obtain a path integral that may be evaluated with
conventional methods, we have introduced a Wiener
regularization. Then, the semiclassical approximation was shown to be
well defined leading to a classical trajectory with jumps at the
endpoints. The action of this trajectory determines the exact quantum
mechanical propagator. Hence, the dominant stationary phase
approximation without fluctuations was shown to become exact for a
spin-$\frac{1}{2}$ system in an arbitrary time-dependent magnetic
field. A non-linear transformation related to the
stereographic projection from the south pole onto the equatorial
plane was found to simplify the explicit determination of the minimal
action trajectory. The method was illustrated by applying it to two
specific models. 

The theory presented has a straightforward extension to spin systems
with quantum numbers $s>\frac{1}{2}$ provided the Hamiltonian remains
of the form (\ref{eq22c}) which is, however, no longer the most
general spin Hamiltonian in this case. For other Hamiltonians, for
$s>\frac{1}{2}$, the dominant stationary phase approximation cannot be
expected to remain exact. An interesting extension of the present work
would be the investigation of the semiclassical dynamics of a spin 
coupled to other degrees of freedom, e.g., boson
modes. With a proper c-number representation of these modes, the
problem can be described as a spin-$\frac{1}{2}$ in a fluctuating
field, and the method presented here can be applied. This will be
studied in future work.

\acknowledgments
The authors would like to thank Joachim Ankerhold, Andreas Lucke, Phil
Pechukas,  Gerhard Stock and Simone Warzel for valuable
discussions. One of us (A~A) is grateful to the Department of
Chemistry of Columbia University, New York, for hospitality during an
extended stay. This work was supported by the Deutsche
For\-schungs\-ge\-mein\-schaft (Bonn) through the Schwer\-punkt\-pro\-gramm 
``Zeit\-ab\-h\"angi\-ge Ph\"anomene und Methoden in Quan\-ten\-sys\-te\-men 
der Phy\-sik und Che\-mie''. Additional support was
provided by the Deutscher Aka\-de\-mi\-scher Aus\-tausch\-dienst (DAAD).




\end{document}